\documentclass[iop]{emulateapj}

\usepackage{graphicx}
\usepackage{apjfonts}
\slugcomment{To appear in ApJ, vol 749, 2012}
\shorttitle{Quiet Sun Internetwork Magnetic Fields}
\shortauthors{Orozco Su\'arez \& Bellot Rubio}

\newcommand{\degree}{\ensuremath{^\circ}\/}

\begin{document}{

\title{Analysis of Quiet-Sun Internetwork Magnetic Fields Based on  
Linear Polarization Signals}

\author{D.\ Orozco Su\'arez\altaffilmark{1,2} and L.R.\ Bellot Rubio\altaffilmark{2}}

\email{d.orozco@nao.ac.jp}

\altaffiltext{1}{National Astronomical Observatory of Japan, 2-21-1 Osawa, Mitaka, Tokyo 181-8588, Japan}

\altaffiltext{2}{Instituto de Astrof\'{\i}sica de Andaluc\'{\i}a (CSIC), Apdo.\ de Correos 3004, 18080 Granada, Spain}

\begin{abstract}
We present results from the analysis of \ion{Fe}{1} 630~nm
measurements of the quiet Sun taken with the spectropolarimeter of the
{\it Hinode} satellite. Two data sets with noise levels of $1.2 \times
10^{-3}$ and $3 \times 10^{-4}$ are employed. We determine the
distribution of field strengths and inclinations by inverting the two
observations with a Milne-Eddington model atmosphere.  The inversions
show a predominance of weak, highly inclined fields. By means of
several tests we conclude that these properties cannot be attributed
to photon noise effects. To obtain the most accurate results, we focus
on the 27.4\% of the pixels in the second data set that have linear
polarization amplitudes larger than 4.5 times the noise level. The
vector magnetic field derived for these pixels is very precise because
both circular and linear polarization signals are used
simultaneously. The inferred field strength, inclination, and filling
factor distributions agree with previous results, supporting the idea
that internetwork fields are weak and very inclined, at least in about
one quarter of the area occupied by the internetwork. These properties
differ from those of network fields.  The average magnetic flux
density and the mean field strength derived from the 27.4\% of the
field of view with clear linear polarization signals are
16.3~Mx~cm$^{-2}$ and 220~G, respectively. The ratio between the
average horizontal and vertical components of the field is
approximately 3.1. The internetwork fields do not follow an isotropic
distribution of orientations.
\end{abstract}


  \section{Introduction}
  \label{Introduction}

Only a small fraction of the solar surface is covered by active
regions at any one time. The rest is occupied by the quiet Sun network
and internetwork (IN). Because of their vast extension, these regions
may contain most of the magnetic flux of the solar surface. It is
therefore important to determine their magnetic properties.

During the last decade, the analysis of Zeeman-sensitive lines has
benefited from advances in solar instrumentation and spatial
resolution. An important contribution has come from the analysis of
the nearly diffraction-limited measurements obtained with the spectropolarimeter
(SP; \citealt{2001ASPC..236...33L}) of the Solar Optical Telescope (SOT;
\citealt{2008SoPh..249..167T, 2008SoPh..249..197S, 
2008SoPh..249..221S,2008SoPh..249..233I}) aboard {\it Hinode}
\citep{2007SoPh..243....3K}. This instrument performs full 
Stokes polarimetry of the \ion{Fe}{1} 630.2~nm lines at high angular
resolution (0\farcs32).  The {\em Hinode}/SP data have led to a better
determination of the magnetic field strength, inclination, and filling
factor distributions in the IN through Milne-Eddington inversions of
the radiative transfer equation \citep{Orozco1,Orozco2}. These results
and the work of \cite{Lites1,Lites2}, \cite{2007A&A...466.1131R}, 
\cite{2008A&A...477..953M}, \cite{2009A&A...502..969B}, and 
\cite{2009A&A...506.1415B} have helped to settle the main properties 
of quiet Sun IN fields: they are weak (of the order of hG) and appear
to be very inclined.

\begin{figure*}[!t]
\centering
\epsscale{1.15}
\plotone{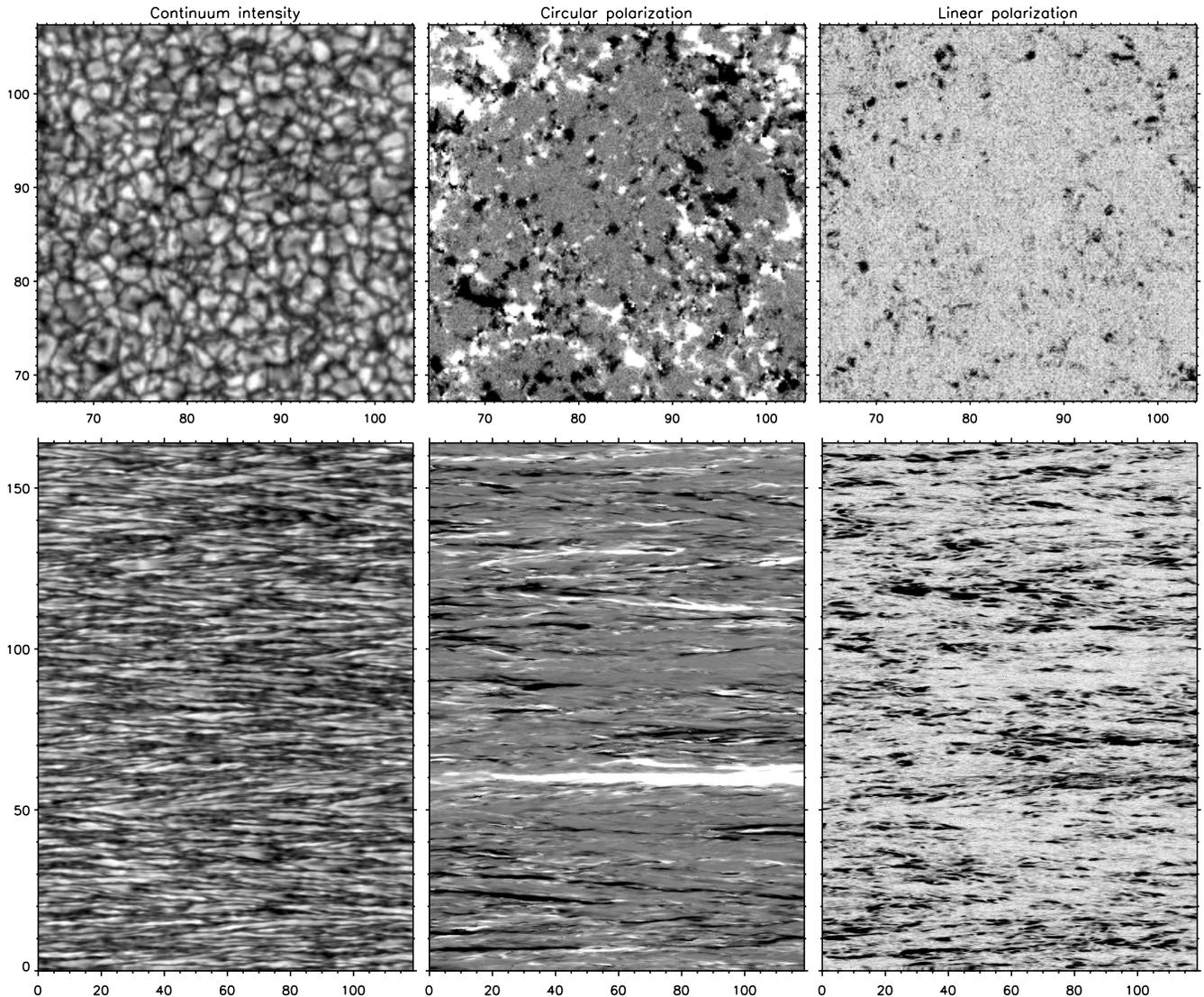}
\caption{{\em Top:} Small IN subfield of 
40\arcsec\/$\times$40\arcsec\ corresponding to data set \#1. From left
to right: continuum intensity, mean degree of circular polarization,
and mean degree of linear polarization. The granulation contrast is
7.44\%. This area contains a supergranular cell of about
30\,000~km$^2$.  Tickmarks are spaced by 1 arcsec. {\em Bottom:} Quiet
Sun region covered by data set \#2. The contrast is 7.3\%. The x-axis
represents time in minutes and the y-axis distance along the slit in
arcsec. In both cases, the polarization maps have been computed as $ [
\sum_{i=N_{\rm b}}^{N_{\rm zc}} V(\lambda_i)/I(\lambda_i) -
\sum_{i=N_{\rm zc}}^{N_{\rm r}} V(\lambda_i)/I(\lambda_i) ] / (N_{\rm
r} - N_{\rm b})$ and $\sum_{i = N_{\rm b}}^{N_{\rm r}}
\sqrt{Q(\lambda_i)^2+U(\lambda_i)^2}/I(\lambda_i)/(N_{\rm r} - N_{\rm
b})$, respectively. $N_{\rm b}$ is the index where the Stokes V blue
lobe of \ion{Fe}{1} 630.25~nm starts, and $N_{\rm r}$ the index where
the red lobe ends. $N_{\rm zc}$ represents the index of the Stokes V
zero crossing. $N_{\rm r} - N_{\rm b} = 22$ is the number of
wavelength points used for the calculations. The circular and linear
polarization maps are saturated at $\pm0.15\%$ of the quiet Sun
continuum intensity.}
\label{fig1}
\end{figure*}

However, there still exists some skepticism about the accuracy of the
magnetic parameters derived from ME inversions of the {\em Hinode}
measurements (e.g.,
\citealt{2009SSRv..144..275D, 2009A&A...506.1415B}). Doubts affect 
particularly the distribution of field inclinations.
\cite{2009ApJ...701.1032A} and \cite{2011A&A...527A..29B}, using 
two different statistical analyses, recently pointed out that the
information contained in the Stokes profiles is not enough to
constrain the magnetic field inclination when the polarization signals
are small and Stokes Q and U are buried in the noise. As a
consequence, the true amount of inclined fields may not be recovered
accurately without clear linear polarization signals. The model
assumptions may also induce large uncertainties in the retrieved
inclinations.

Always present in real observations, photon noise hides the weakest
polarization signals and distorts the larger ones. There are two ways
to partly circumvent this problem: to reduce the noise itself and/or
to increase the signal. Both strategies lead to higher signal-to-noise
ratios (S/N), allowing for a more precise discrimination between the
magnetic field strength, field inclination, and filling factor in the
quiet photosphere. The noise can be reduced by using larger telescopes
or longer effective exposure times; the signals are enhanced by
improving the spatial resolution, since this minimizes the
cancellation of opposite-polarity fields.

Here we update our results from the inversion of {\it Hinode}/SP data
\citep{Orozco1,Orozco2,Orozcotesis} by analyzing new measurements with
significantly better S/N of up to 3500. In Section~2 we describe the
observations, study the amplitudes of the observed polarization
signals, and give an account of the Milne-Eddington inversion strategy
we follow.  In Section~3 the results of the inversion are presented.
Section~4 discusses the effects of noise on the inferences, taking
advantage of the availability of two data sets with different S/N. In
Section~5 we determine very precise distributions of field strength
and field inclination for pixels that show {\em linear} polarization
signals well above the noise level. The magnetic flux density, the
average field strength, and the vertical and horizontal components of
the field are calculated and compared with previous observational
determinations and magneto-convection simulations in Section~6. To
highlight the distinct character of IN fields, Section~7 describes the
magnetic properties of the network as deduced from the
inversion. Finally, we discuss our results in Section~8.

\section{Observations, noise analysis, and inversion of Stokes profiles}
\label{Observations}

The observations analyzed in this paper come from two different data
sets taken at disk center. Hereafter they will be referred to as
normal map observations (set \#1) and high S/N time series (set
\#2). \cite{Lites1, Lites2}, \cite{Orozco1,Orozco2}, \cite{2009ApJ...701.1032A}, 
\cite{2011A&A...527A..29B}, and \cite{2011ApJ...731..125A} have also used
these measurements. Their main parameters are summarized in
Table~\ref{Table1}.

In observation \#1, the spectrograph slit (of width 0.16\arcsec\/) was
moved across the solar surface in steps of 0\farcs1476 to measure the
four Stokes profiles of the \ion{Fe}{1} 630~nm lines with a spectral
sampling of 2.15~pm~pixel$^{-1}$ and a exposure time of 4.8~s. For
data set \#2, the slit was kept fixed at the same spatial location and
the Stokes spectra were recorded with an exposure time of 9.6~s. The
completion of data set \#1 took about 3 hours while the time series of
data set \#2 covers one hour and 51 minutes.

The effective exposure time of observation \#2 was increased by
averaging seven consecutive 9.6~s measurements. This allowed a final
exposure time of 67.2~s to be reached, which corresponds to a S/N gain
by a factor of about 3.8 with respect to data set \#1. The averaging
decreased the spatial resolution of the observations to some degree,
but the granulation pattern is still perfectly visible due to the
longer lifetime of photospheric convective structures (about 5~min,
see \citealt{1998ASPC..154..345T}) and the excellent stabilization
system of {\it Hinode}
\citep{2008SoPh..249..221S}. Indeed, the rms continuum contrast of the
granulation in the high S/N series is only 0.14\% smaller than in the
normal map. This variation is real, since the SOT focus position was
optimized for the SP observations in the two cases.

\begin{table}[!t]
\centering
\caption{Log of the observations}
\label{Table1}
\begin{tabular}{@{}lcc@{}}
\hline
\, & {\scshape Data set \#1} & {\scshape Data set \#2} \\\, & {\scshape (Normal map)} & {\scshape (High S/N time series)} \\ \hline Date &
2007 March 10 & 2007 February 27 \\ Start time (UT) & 11:37:37 & 00:20:00 \\
FOV & 302\arcsec\/$\times$162\arcsec\/ &
302\arcsec\/$\times$0\farcs16 \\ \,\,
(Pixels) & 1024$\times$2048 & 1024$\times$727 \\ Exposure time & 4.8~s &
67.2~s \\ Stokes V noise ($\sigma_{\rm V}$) & $1.1\times10^{-3} \, I_{\rm QS}$ &
$2.9\times10^{-4} \, I_{\rm QS}$ \\ Stokes Q,U noise ($\sigma_{\rm Q,U}$)&
$1.2\times10^{-3} \, I_{\rm QS}$ & $3\times10^{-4} \, I_{\rm QS}$ \\
\hline
\end{tabular}
\end{table}

The polarization noise levels ($\sigma$) are shown in Table~\ref{Table1}. 
They were obtained as the mean value of the standard deviation of the
corresponding Stokes signals in continuum wavelengths. Before
evaluating the noise, the data were corrected for dark current,
flat-field, and instrumental crosstalk. The whole process was
accomplished using the routine \texttt{sp\_prep.pro} included in
SolarSoft. The Stokes profiles were normalized to the average quiet
Sun continuum intensity, $I_{\rm QS}$, computed using all pixels from
each data set.

\begin{figure}[!t]
\centering
\epsscale{1}
\plotone{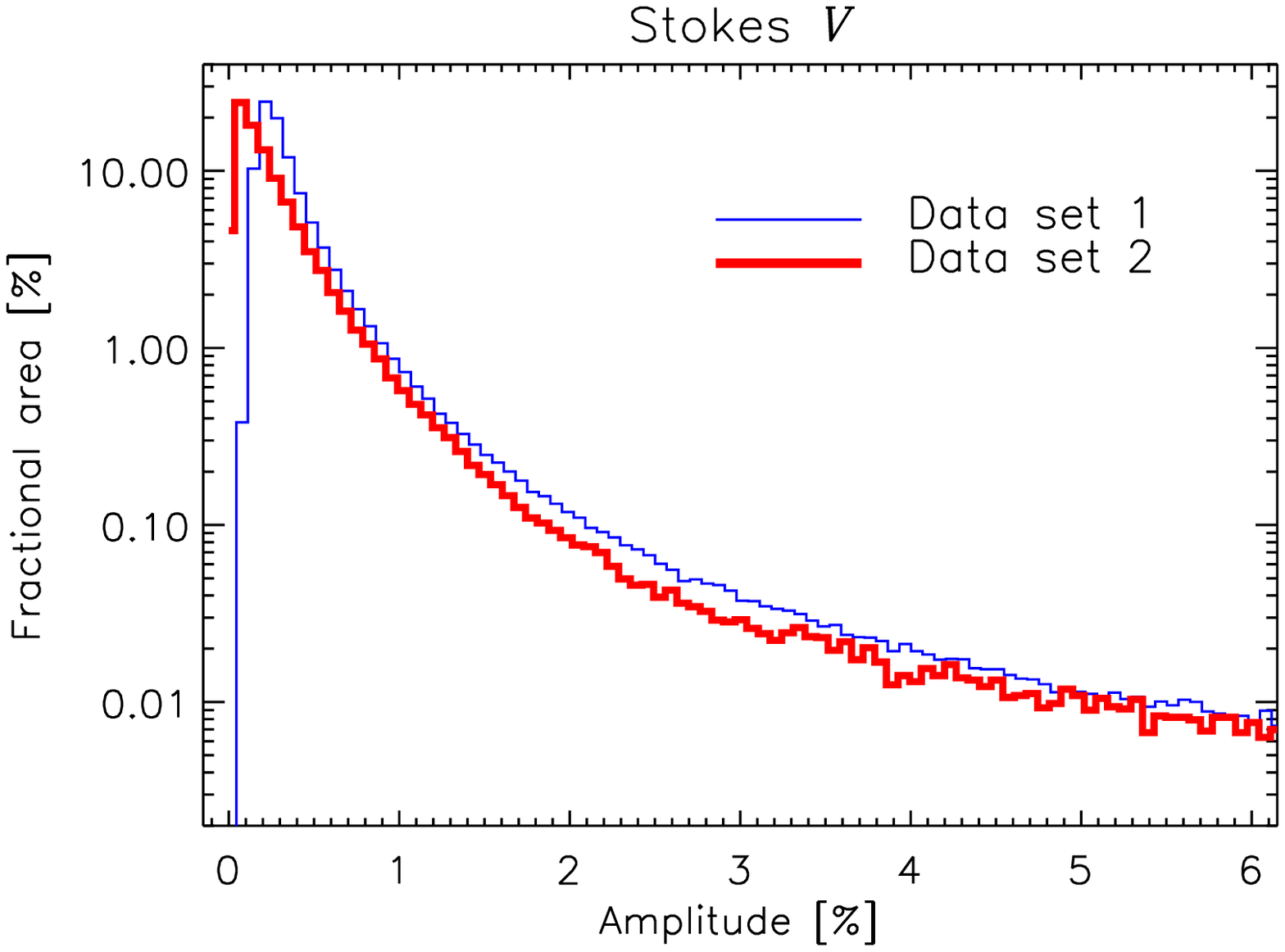}
\plotone{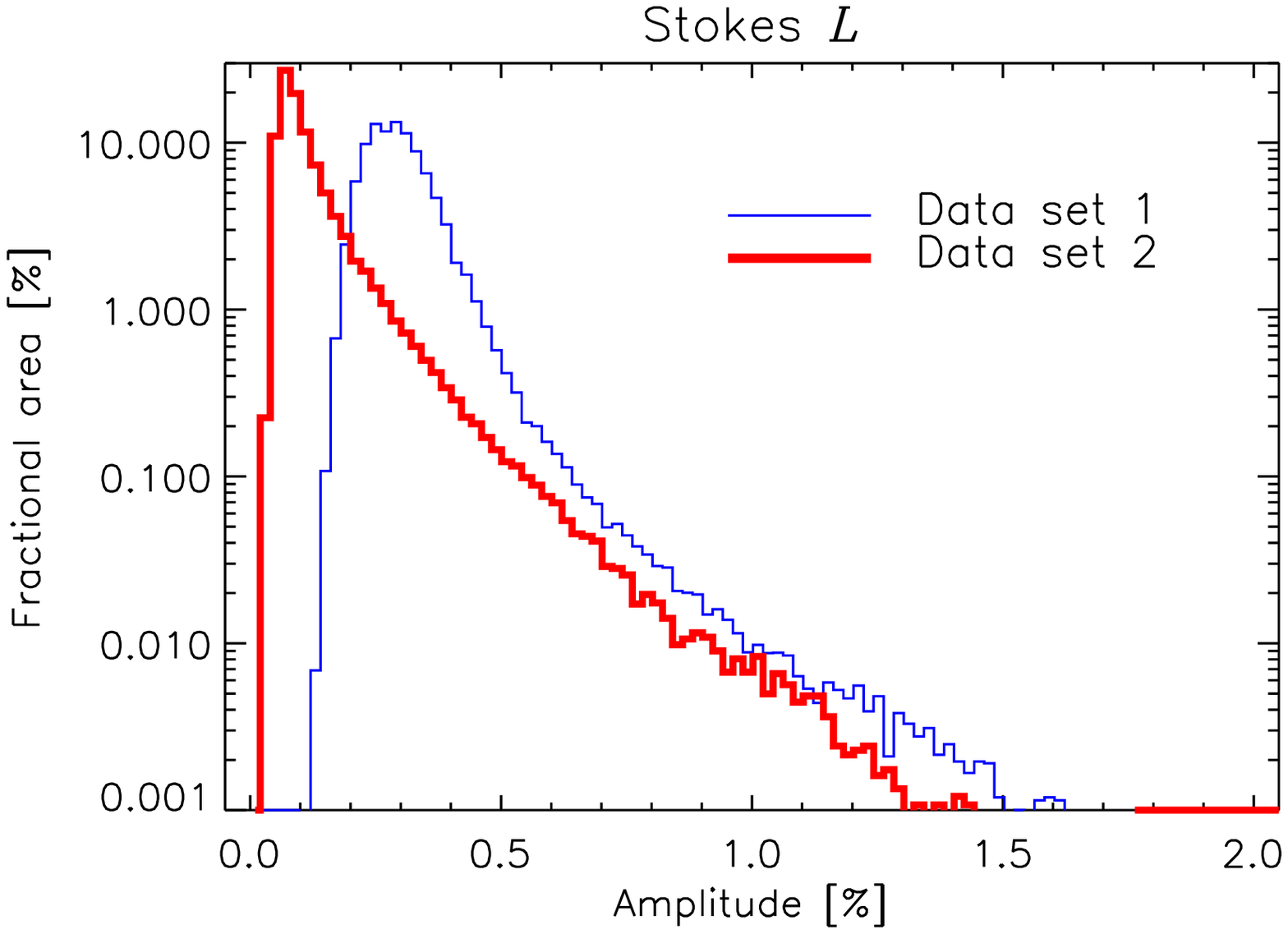}
\plotone{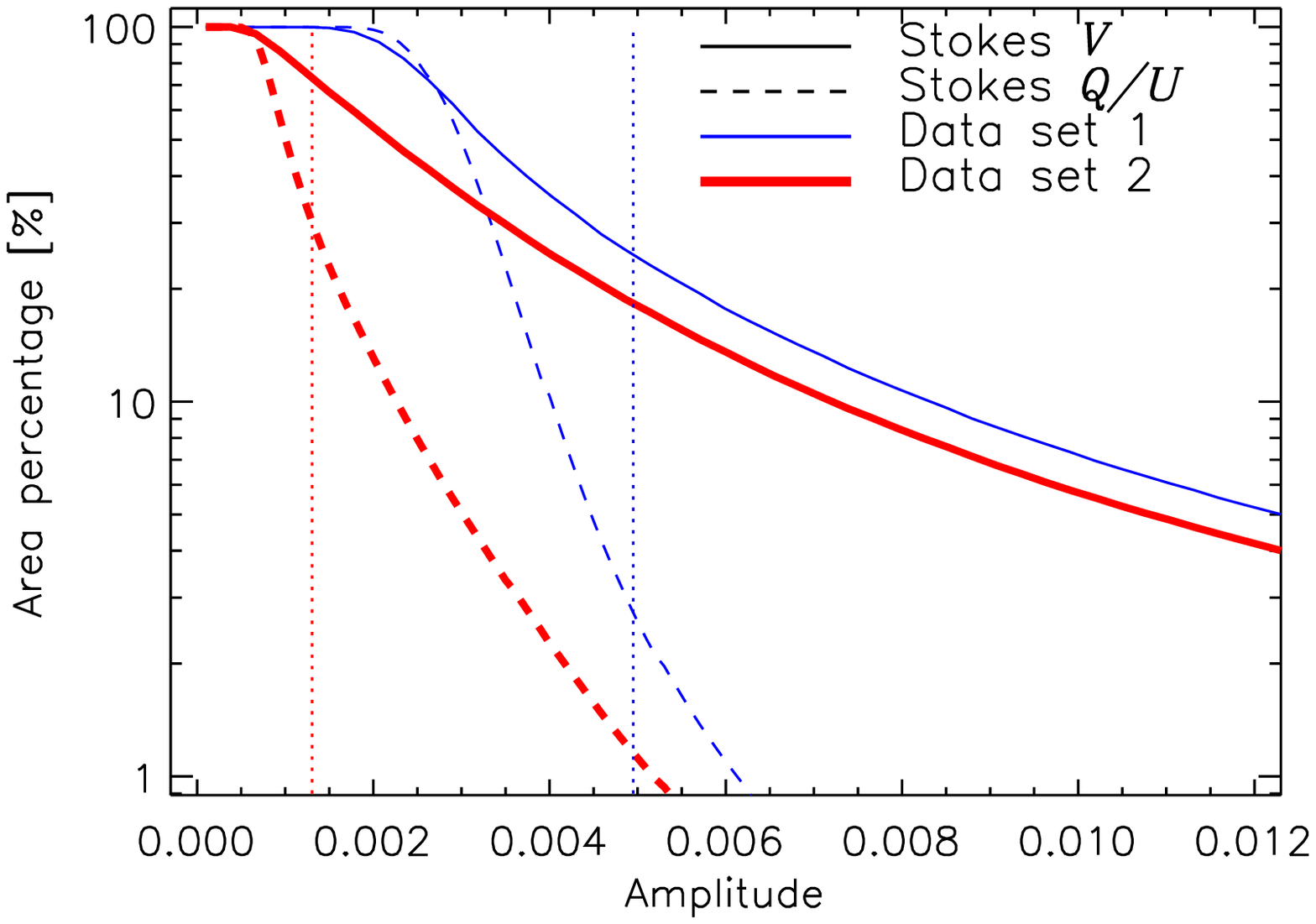}
\caption{{\em Top:} Amplitudes of the observed Stokes 
V profiles, in units of $I_{\rm QS}$. The thin and thick lines indicate
the normal map and the high S/N time series, respectively. {\em
Middle:} Same, for Stokes Q and U (the one with the largest amplitude
is selected).  {\em Bottom:} Surface area occupied by signals with
Stokes V or Stokes Q or U amplitudes larger than a given value. Vertical
lines mark the 4.5$\sigma$ noise thresholds for the two data sets
analyzed in this paper.}
\label{fig2}
\end{figure}

Figure \ref{fig1} displays maps of the continuum intensity and mean
circular and linear polarization degrees for data sets \#1 and
\#2. The different nature of the two observations is apparent. 
The normal map, as a ``snapshot'' of the solar surface, reveals the
spatial distribution of convective cells and magnetic flux
concentrations in the quiet Sun. By contrast, data set \#2 shows the
evolution of these structures.  For instance, granules (intergranules)
are seen as bright (dark) horizontal streaks. Flux concentrations also
produce horizontal streaks which last longer in circular
polarization. Note that data set
\#2 exhibits many more linear polarization patches than the normal map
because of the lower noise level.

Figure \ref{fig2} shows histograms of the circular (Stokes V) and
linear (Q, U) polarization amplitudes of \ion{Fe}{1} 630.25~nm (top
and middle panels, respectively) in the two data sets. They
demonstrate the large occurrence of weak polarization signals. The
distributions are similar for the high S/N time series and the normal
map, although some differences exist. For example, the normal map
histograms peak at larger amplitudes. In both data sets, however, the
peaks are close to the corresponding noise levels: the Stokes V
distributions reach their maxima at about 1.95$\sigma_{\rm V}$ and
2.35$\sigma_{\rm V}$, respectively, while the linear polarization
histograms peak at 2.42$\sigma_{\rm Q,U}$ and 2.33$\sigma_{\rm Q,U}$.

To avoid polarization signals that are highly contaminated by noise,
we only consider pixels with Stokes Q, U {\em or} V amplitudes larger
than 4.5$\sigma_{\rm V}$. This criterion is very restrictive: the
probability that one pixel showing pure noise in all three Stokes
parameters be included in the analysis because one of the signals
exceed by chance the 4.5$\sigma_{\rm V}$ level is only 0.19\%. This
number results from the multiplication of three factors: the
probability of $7 \times 10^{-6}$ that a single measurement of a zero
signal exceeds $\pm 4.5 \sigma_{\rm V}$ when the noise follows a
normal distribution, the three Stokes parameters that we consider, and
the 90 wavelength points used to compute the amplitude of the
polarization signals. With such a low probability, we can be sure that
the inverted profiles are real and not due to noise.

\begin{figure*}[!t]
\centering
\epsscale{1.15}
\plotone{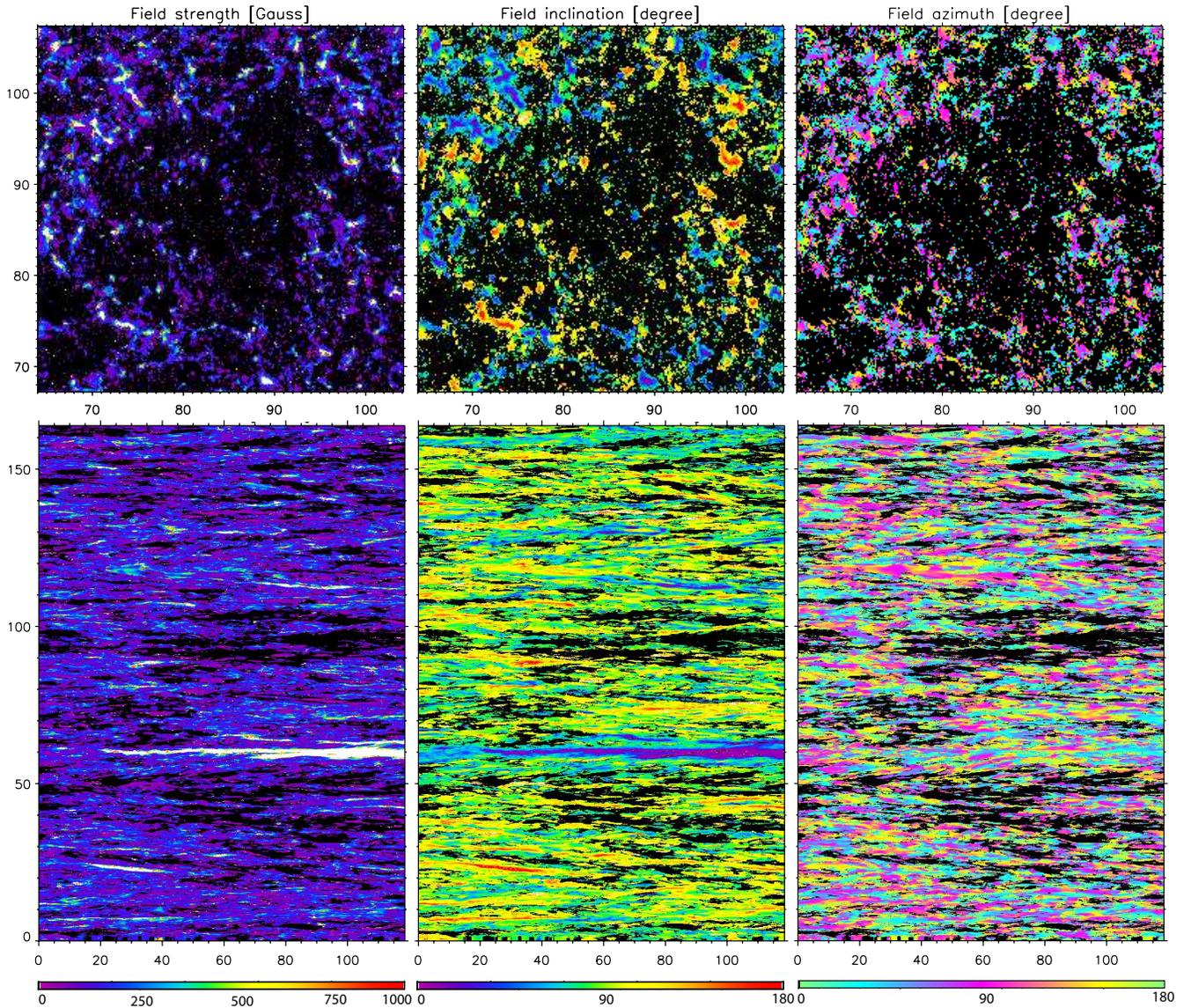}
\caption{Magnetic field parameters resulting from the inversion of 
the normal map (top row) and the high S/N time series (bottom
row). Shown from left to right are magnetic field strengths,
inclinations, and azimuths. Pixels in black did not meet the selection
criterion and were not inverted. The field strength map is
saturated at 1~kG so the pixels in white correspond to fields $\geq$
1~kG. The FOVs are the same as in Figure~1.  Tickmarks are spaced by
1 arcsec in the normal map. In the high S/N time series the x-axis
represents time in minutes and the y-axis distance along the slit in
arcsec. }
\label{fig3}
\end{figure*}

\begin{figure*}[!t]
\centering
\plotone{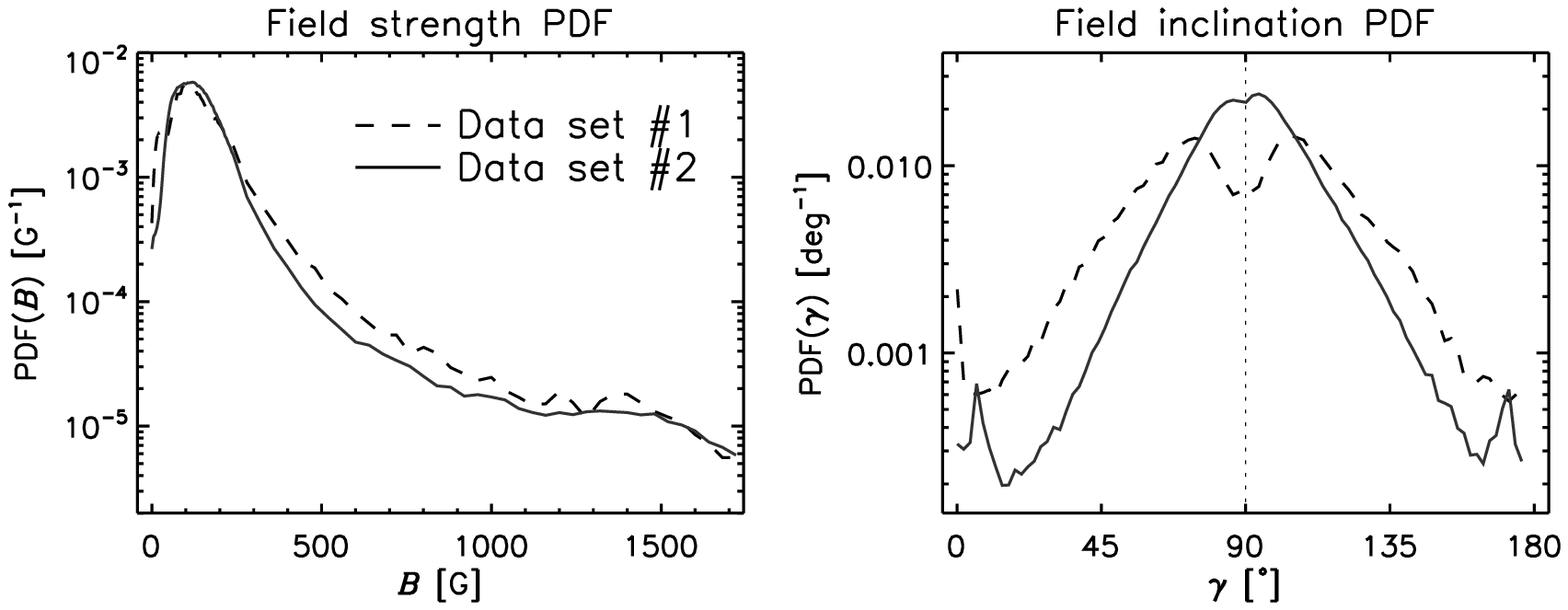}
\caption{Magnetic field strength (left) and inclination (right) 
probability density functions for IN regions resulting from the
inversion of data sets \#1 and \#2 (dashed and solid lines,
respectively). The PDF for the magnetic field strength corresponding
to data set \#2 resamples quite well the one corresponding to the
normal map, despite the very different S/N. These PDFs are derived
from the analysis of pixels with Stokes Q, U or V amplitudes larger
than 4.5 times the noise level. The fraction of the FOV they cover 
is 26.8\% in the normal map and 72.7\% in the high S/N time series. 
Of these pixels, 7.9\% and 37.7\% show linear polarization signals above
4.5$\sigma_{\rm Q,U}$.}
\label{fig4}
\end{figure*}

The bottom panel of Figure \ref{fig2} shows the fraction of the FOV
with Stokes V signals above a given amplitude. A similar curve is
provided for the linear polarization signals (Stokes Q or U). The
vertical lines indicate 4.5 times the noise level in Stokes V for data
sets \#1 and \#2. Only 26.0\% of the normal map area exhibit Stokes V
amplitudes above our $4.5\sigma_{\rm V}$ noise threshold. This
fraction increases to 70.1\% in the high S/N time series. The area
with Stokes Q or U signals larger than 4.5$\sigma_{\rm Q,U}$ is
smaller, only 2.1\% and 27.4\%, respectively. In summary, the fraction
of pixels fulfilling the selection criterion (i.e., Stokes Q, U, or V
amplitudes larger than 4.5 times their noise levels) is 26.8\% and
72.7\% in data sets
\#1 and \#2 (corresponding to about 0.7 and 1.6~Mpx). For comparison,
62.6\% and 85.7\% of the FOV covered by the two observations show
Stokes V signals above $3\sigma_{\rm V}$ (cf.\ the 92.6\% of
\citealt{2008A&A...477..953M} and the 80-90\% reported by
\citealt{khomenko}, both from ground-based observations).

In the normal map the selection of IN areas has been done manually to
avoid the strong flux concentrations of the network (see
\citealt{Orozcotesis} for details). In the high S/N time series we
have just removed the strong magnetic feature of positive polarity
visible  at around 1/3rd of the slit
($y=60$\arcsec). 

The observations are analyzed using least-squares inversions based on
a one-component, horizontally homogeneous ME atmosphere, as described
by \cite{2007ApJ...662L..31O,Orozco1,Orozco2}.  We include a local 
non-magnetic component to account for the reduction of the
polarization signals induced by the instrument's Point Spread Function
(PSF). This component was called ``local'' stray-light contamination
in \cite{2007ApJ...662L..31O}. Here we use a different name to better
reflect the nature of this contribution and distinguish it from real
stray light, i.e., light that enters the detector because of unwanted
reflections in the mirror surfaces, supporting structures, etc. For
the normal map, the local non-magnetic contribution is evaluated
by averaging the Stokes I profile within a 1\arcsec-wide box centered
on the pixel. For the high S/N time series we take the non-magnetic
component as the average Stokes I profile along 1\arcsec\/ of the SP
slit centered on the pixel. The reason is that we cannot perform a
two-dimensional average since the data is one-dimensional. With this
approximation we avoid using measurements acquired more than a minute
apart, although the non-magnetic profile may not appropriately account
for the effects of the PSF. Further information about how the PSF
changes the polarization signals are given in
\cite{2007ApJ...662L..31O}\footnote{The instrumental PSF considered by
these authors included telescope diffraction, CCD pixelation, and SP
slit width. Other optical aberrations and stray light were not taken
into account.} and
\citet{2010AN....331..558D}. The implications of using a local stray 
light have been analyzed by \cite{2011ApJ...731..125A}. In practice,
our approach is equivalent to a two-component inversion in which the
non-magnetic atmosphere has only one free parameter, $(1-f)$, with $f$
the magnetic filling factor. The two \ion{Fe}{1} lines were inverted simultaneously with the MILOS code
\citep{2007A&A...462.1137O,2010A&A...518A...3O}. The use of two lines, 
as opposed to only one (e.g., \citealt{2009A&A...506.1415B}), reduces
the influence of noise and leads to more accurate results
\citep{2010A&A...518A...3O}.

\section{Inversion results} 

Maps of the field strengths, inclinations, and azimuths retrieved from
the two data sets are shown in Figure~\ref{fig3} (see also
\citealt{Orozco1,Orozco2} and \citealt{Orozcotesis}).  Pixels in black were not
inverted because their Stokes Q, U, or V amplitudes do not exceed
4.5$\sigma_{\rm V}$. Qualitatively, the inversions of the two data
sets give the same results. Most of the fields are weak (of the order
of hG), with the stronger field concentrations located in
intergranular lanes. The fields are very inclined, especially in
granules.  The azimuth is much better recovered in the high S/N time
series due to the lower noise of the linear polarization profiles.

Figure~\ref{fig4} shows the probability density function (PDF;
\citealt{2003A&A...406.1083S}) of magnetic field strengths and field
inclinations in the IN for the normal map and the high S/N time series
(dashed and solid lines, respectively). We recall that they were
obtained from the analysis of Stokes profiles whose amplitudes exceed
a polarization threshold of $4.5\sigma_{\rm V}$.  The S/N of data set
\#2 is about 3.8 times better than that of the normal map. However, the 
peaks of the field strength distributions are located at nearly 
the same position ($\sim$~90~G) and have very similar widths.  A
closer inspection shows that the maximum of the field strength
distribution for the high S/N time series occurs at around 100~G,
i.e., at slightly stronger fields. These PDFs seem to indicate that
the IN consists of hG flux concentrations. In general, the field
strength PDF does not change when using better S/N data, suggesting
that the inversion results are reliable and not affected by the noise
of the profiles.

The inclination PDF for the normal map has a maximum at 90\degree\/
and decreases toward vertical orientations. Near 0\degree\/ and
180\degree\/ the PDF increases again.  The distribution of field
inclinations derived from the high S/N data is very similar, with some
minor differences: first, the amount of nearly horizontal fields
increases with respect to that in the normal map, indicating that the
smallest polarization signals (not considered before) are also
associated with highly inclined fields; second, the PDF is narrower,
so that the fields are less abundant as they become more vertical.
Overall, the two distributions point to large IN field inclinations.

\section{Reliability of the inferred magnetic field distributions}

\begin{figure*}[!t]
\begin{center}
\plotone{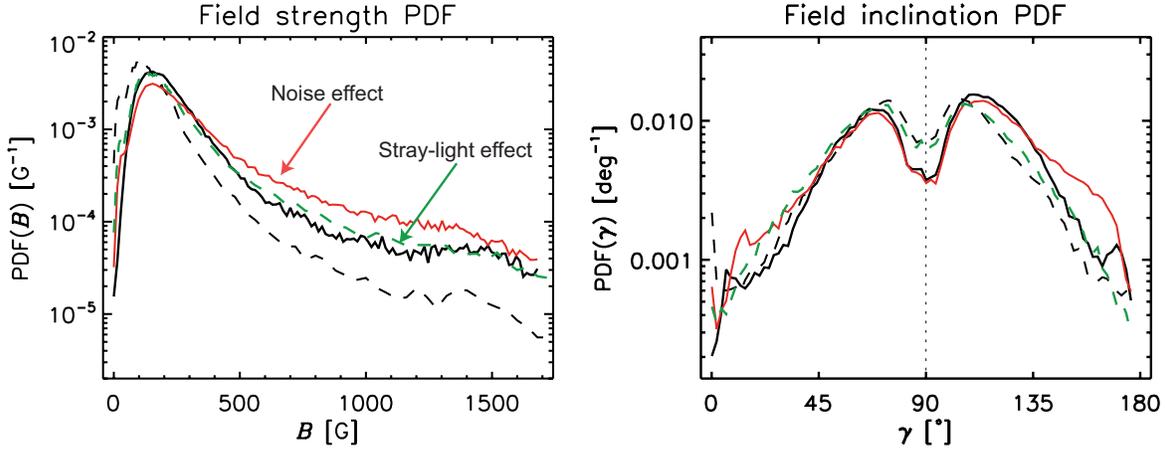}
\end{center} 
\caption{Magnetic field strength (left) and field inclination (right) 
distributions for IN regions of data sets \#1 and \#2 (dashed and
solid lines, respectively). They have been determined from pixels
whose Stokes Q, U or V amplitudes exceed the same absolute threshold
of 0.5\% the quiet Sun continuum intensity (black curves).  For the
normal map, this amplitude represents $4.5\sigma_{\rm V}$ (and the
corresponding PDFs are identical to those displayed in
Figure~\ref{fig4}). For the high S/N data, an amplitude of $0.5\%
\, I_{\rm QS}$ is 16.4 times {\em larger} than the noise. The percentage
of pixels fulfilling the selection criterion is 26.8\% in the normal
map and 16.4\% in the high S/N data. Of these pixels, 7.9\% and 6.8\%
show linear polarization signals above the noise threshold,
respectively. The solid red lines stand for the inversion {\em of the
same high S/N profiles} once their noise has been artificially
increased to a level comparable to that found in data set \#1. The
percentage of analyzed pixels is then 16.4\%, of which 6.8\% show
linear polarization signals above 0.5\%~I$_{\rm QS}$. Finally,
thegreen lines represent the results of the inversion of data set \#1
calculating the non-magnetic component as in data set
\#2.}
\label{fig5}
\end{figure*}

The polarization signals we measure in the internetwork are tiny
compared to those found in active regions. As a consequence, a careful
treatment of the noise is important to interpret them correctly. The
first obvious choice to minimize the effects of noise was to set a
rather conservative threshold on the amplitude of the polarization
signals, which had to exceed 4.5 times the noise level to be included
in the analysis. We used this criterion to identify and avoid the
noisier signals that cannot be inverted reliably.

The threshold of $4.5\sigma_{\rm V}$ was chosen taking into
consideration the results presented by \cite{2007ApJ...662L..31O} and
\cite{Orozcotesis}. There, {\it Hinode}/SP normal map observations
were simulated with the help of 3D magnetoconvection models to
determine whether it is possible to recover the distribution of
magnetic field strengths and inclinations with a S/N of 1000 and a
threshold of $4.5\sigma_{\rm V}$. The results showed that the field
strength and the field inclination can be recovered with mean errors
of less than 100~G and 10\degree, respectively. For very weak fields
($B \sim 100$~G), the rms uncertainty of the inclination is smaller
that 30\degree, sufficient to distinguish between highly inclined and
vertical fields. The errors in field strength and inclination are
expected to be smaller for measurements with lower noise levels, such
as the high S/N time series.

However, the influence of noise is still of concern. 
\cite{2009ApJ...701.1032A} analyzed a small region of the
normal map using Bayesian techniques and Milne-Eddington
inversions. His results suggest that while the magnetic field strength
tends to be well recovered, it is in general not possible to constrain
the field inclination when the signals are very weak and Stokes Q or U
do not stand out prominently above the noise.  A possible consequence
of the limited information carried by the Stokes vector when Q and U
are buried in the noise is an overabundance of horizontal fields. This
is because inversion algorithms try to fit the noise of the linear
polarization profiles
\citep{2003A&A...408.1115K}.

Also \cite{2011A&A...527A..29B} have recently argued that the noise
present in Stokes Q and U can potentially distort the Milne-Eddington
inferences. These authors showed how in some cases inclined fields may
be obtained from purely vertical weak magnetic fields. This happens
when the noise in Stokes Q and U is interpreted as a real signal.  In
the weak field regime, producing linear polarization in \ion{Fe}{1}
630~nm at the level of the noise requires large transverse fields.
Thus, the noise may be compatible with highly inclined fields.  An
additional effect is that the recovered fields are stronger than the
real ones. \cite{2011A&A...527A..29B} inverted the two data sets used
in this paper (but only one spectral line, not the two) and detected a
small shift of the peak of the field strength PDF toward weaker fields
for the data with better S/N, which they interpreted as evidence that
noise was playing an important role.

There are reasons to believe that the field strength and field
inclination distributions we have derived are reliable and not
significantly affected by the noise, even if most of the pixels do not
show significant Stokes Q or U signals. The following arguments
support this conclusion:
\begin{itemize}
\item[-] If noise were affecting the inversions, the fields obtained 
from the normal map (the noisier one) would be stronger and more
inclined than those from the high S/N data, as claimed by
\cite{2011A&A...527A..29B}. Our distributions indeed show a small
shift when the S/N is improved, but in the opposite direction, 
i.e., toward stronger fields (see Figure~\ref{fig4}). 
\item[-] Although only 27.4\% of the pixels in the high S/N time
series have linear polarization signals above $4.5\sigma_{\rm Q,U}$,
the maps of magnetic field inclination and particularly the azimuth
derived from the inversion are dominated by structures which vary
smoothly both in time and in space (Figure~\ref{fig3}).  These
structures are real and not caused by the noise. Therefore, the
smallest linear polarization signals below $4.5\sigma_{\rm Q,U}$ 
also provide information to constrain the field azimuth and
inclination.
\item[-] In the absence of clear Stokes Q and U signals it may
still be possible to gather information about the field inclination:
first, the noise in Q and U sets limits on it, and second, in the weak
field regime Stokes I has greater sensitivity to the inclination than
Stokes Q, U, or V \citep{2010AN....331..558D}.
\end{itemize}

At this point, a solid demonstration that noise is not affecting the
determination of field strengths and inclinations would come from the
similarity of the results derived from pixels which show the same type
of polarization signals but have very different noise levels. Thus, in
Figure~\ref{fig5} we compare the PDFs calculated from pixels with
Stokes Q, U, {\em or} V signals larger than 0.5\%~I$_{\rm QS}$ in both
data sets.  For the normal map (black dashed line), this amplitude
threshold represents 4.5 times the noise level, but for the high S/N
observations (black solid line) the same absolute threshold
corresponds to signals 16 times {\em larger} than the noise.
Unexpectedly, we observe some differences in the shapes of the
resulting PDFs. For example, the peak of the field strength
distribution obtained from the high S/N measurements is slightly
shifted toward stronger fields compared to the normal map. This
implies a smaller abundance of weak fields and a larger fraction of
strong fields.  Also, the two inclination PDFs show a dip at around
90\degree\/, but less pronounced in the case of the high S/N
measurements.

These differences may reflect a better determination of the magnetic
field parameters from the less noisy signals.  Alternatively, they may
indicate intrinsic differences between the two data sets or problems
with the inversion scheme. To identify the actual origin of the differences,
we have carried out two additional tests. In the first one, we have
artificially increased the noise of data set \#2 to the level of the
normal map.  Then we have inverted the pixels with Stokes Q, U, or V
signals above 4.5 times the new noise level. Since this sample of
profiles has the same noise as data set \#1, one should expect PDFs
similar to those obtained from the normal map (black dashed line in
Figure~\ref{fig5}). The differences between the new PDFs and the black
solid lines should only be ascribed to the increased noise. The
outcome of this exercise is shown in Figure~\ref{fig5} with red solid
lines. As can be seen, the new PDFs are closer to the ones derived
from the high S/N measurements (black solid line) than to the normal
map PDFs. Compared with the original high S/N measurements, the
abundance of weak fields ($0 < B < 300$~G) decreases and the fraction
of strong fields ($B > 300$~G) increases, but the peak of the PDF
remains at the same position. The fraction of inclined fields is also
very similar, with slightly less vertical fields. These results
indicate that {\em noise is not producing the differences between the
black curves of Figure~\ref{fig5}}, since the inversion of data set
\#2 with enhanced noise does not reproduce the normal map PDFs. 
Thus, the cause of the differences must be found elsewhere. Another
interesting result is the following. From the comparison of the PDFs
derived from data set \#2 with varying noise levels (black and red
solid curves), we do not confirm the conclusion of
\cite{2011A&A...527A..29B} that an increase of the noise in Stokes Q
and U shifts the PDFs to stronger and more horizontal fields. The red
curves, much more affected by noise than the black curves but
otherwise coming from exactly the same profiles, do not show any of
the two effects. This is probably due to the fact that we analyze real
observations where the fields are not completely vertical, contrary to
the situation modeled by \cite{2011A&A...527A..29B}.

In the second test we examine the influence of the local non-magnetic
contribution, which is calculated differently in the two data sets.
To that end, we have repeated the inversion of the normal map using a
non-magnetic contribution obtained as in the case of the high S/N time
series, i.e., by averaging the Stokes I profiles over 1\arcsec\/ along
the slit.  The results, depicted in Figure~\ref{fig5} with green dashed
lines, show that the fraction of weak fields (below 300~G) decreases
and the amount of strong fields increases with respect to the dashed
black curves. The PDF peak shifts to stronger fields. Remarkably, the
field strength PDF coincides with that calculated from the pixels in
the high S/N time series that show Stokes Q, U, or V amplitudes
larger than 0.5\%~I$_{\rm QS}$. The inclination PDF does not change
much. Overall, this test suggests that (part of) the differences
between the normal map and the high S/N time series observed in
Figure~\ref{fig5} may be the result of the different treatment of the
non-magnetic component in the two data sets.

\section{Analysis based on linear polarization signals}

\begin{figure}[!t]
\begin{center}
\epsscale{1.2}
\plotone{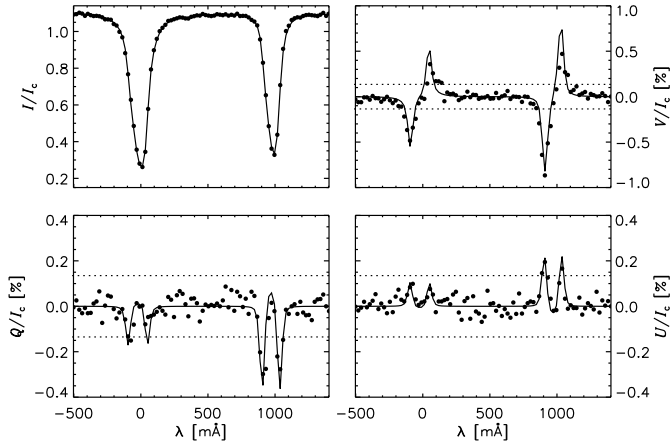}
\end{center} 
\caption{Observed (dots) and best-fit (solid) Stokes profiles corresponding to a 
pixel with linear polarization signals above 4.5 times the noise level. The field strength, inclination,
azimuth, and filling factor retrieved for this pixel are 193~G,
104\degree, 76 \degree, and 33\%. The horizontal lines mark the 4.5$\sigma$
threshold.}
\label{fig6}
\end{figure}

To obtain the most accurate results, in the rest of the paper we will
focus on the analysis of pixels from the high S/N time series showing
Stokes Q or U signals above $4.5\sigma_{\rm Q,U}$. With Q and U
clearly above the noise, a very precise determination of the field
inclination can be made because the inversion code uses the maximum
amount of information possible
\citep{2009ApJ...701.1032A,2011A&A...527A..29B}. The field strength 
is also obtained more accurately, due to the different dependence of
the linear and circular polarization on B in the weak field regime.

To illustrate how these signals look like, Figure~\ref{fig6} shows the
Stokes profiles of a typical IN pixel whose linear polarization is
just above the 4.5$\sigma_{\rm Q,U}$ level. We also include the
corresponding best-fit profiles. The fit is quite successful even
though the signal is close to the threshold and Stokes V is slightly
asymmetric.  Note that the linear polarization signals are above the
noise threshold (horizontal lines) at several wavelength positions,
not just one, which allows the inversion code to determine the field
inclination and the azimuth accurately {\em because the whole
polarization profiles are recognizable}. In this case, the inversion
returns a field strength of 193~G, an inclination of 104\degree, an
azimuth of 76\degree, and a filling factor of 33\%.

\begin{figure*}[!t]
\begin{center}
\plotone{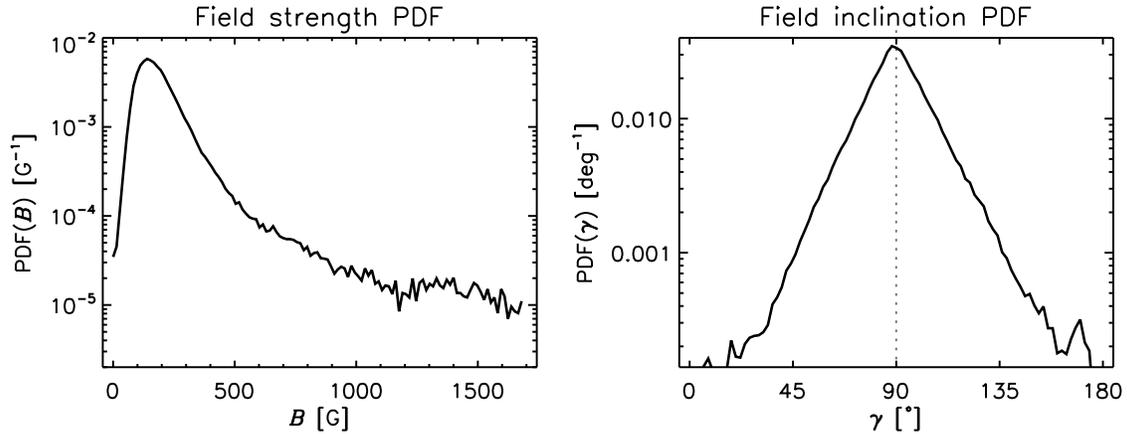}
\end{center} 
\caption{High-precision magnetic field strength and inclination 
distributions for IN regions derived from data set \#2. The PDFs are
based on the inversion of pixels whose Stokes Q or U amplitudes exceed
4.5$\sigma_{\rm Q,U}$. They represent 27.4\% of the surface area
covered by the IN. About 90.3\% of those pixels show circular
polarization signals above the noise threshold.}
\label{fig7}
\end{figure*}

An analysis restricted to pixels showing linear polarization signals
is very reliable because the effects of noise are largely suppressed.
However, the price to pay is a reduced coverage of the internetwork
(only 27.4\% of the IN surface area) and a bias towards inclined
and/or strong fields, since those are the fields that more easily
produce linear polarization signals. We do not know exactly the amount
of bias, but this possibility should be kept in mind. In any case, the
data we are considering represent more than 1/4th of the internetwork,
which is non negligible. Of all the pixels selected for analysis,
90.3\% also show Stokes V amplitudes above $4.5\sigma_{\rm V}$.

Figure~\ref{fig7} shows our most precise determination of the field
strength and field inclination distributions in the IN based on pixels
of the high S/N time series with Stokes Q or U amplitudes exceeding
$4.5\sigma_{\rm Q,U}$. These PDFs confirm the main results derived
from the analysis of all the pixels that met the inversion threshold,
namely that the IN fields are weak and very inclined.  It is
remarkable that the maximum of the field strength PDF remains at more
or less the same position, with only a small shift toward stronger
fields (about 130~G). The inclination PDF is narrower than the ones
displayed in Figure~\ref{fig4}. Very likely, this is due to selection
effects: the more vertical fields are excluded from the analysis
because they produce smaller Stokes Q and U signals.  Therefore, the
amount of vertical fields can be expected to decrease with respect to
that found when analyzing pixels with Q, U, {\em or} V above 4.5 times
the noise level.

Figure ~\ref{fig8} shows the filling factors inferred from the
inversion of the profiles in the high S/N time series with Q or U
signals above 4.5 times the noise level. As can be seen, the maximum
of the distribution occurs at around 0.25, with a tail extending
toward filling factors of up to 0.4--0.6. Such values are
significantly larger than those obtained from ground-based
observations, and imply that some of the IN fields are almost resolved
by the {\it Hinode} SP. This seems plausible in view of the recent
detection of resolved network flux tubes by
\cite{2010ApJ...723L.164L}, using data taken with the IMaX
magnetograph \citep{2011SoPh..268...57M} aboard the SUNRISE
balloon-borne solar observatory \citep{2010ApJ...723L.127S}.  The
resolution of the IMaX measurements is 0\farcs18, nearly twice as good
as that provided by the {\it Hinode} SP.

\begin{figure}[!t]
\begin{center}
\epsscale{1.1}
\plotone{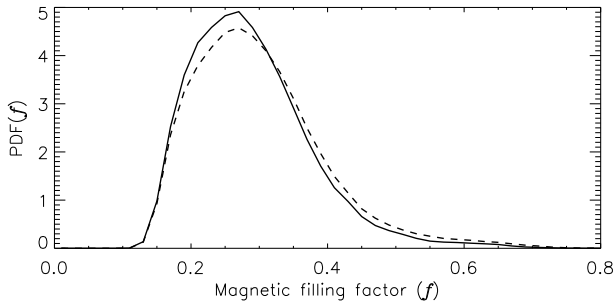}
\end{center} 
\caption{
Distribution of magnetic filling factors retrieved from the inversion
of the high S/N time series including only pixels whose Stokes Q or U
amplitudes exceed 4.5$\sigma_{\rm Q,U}$ (solid). For comparison,
we also represent the distribution corresponding to all pixels whose
Stokes Q, U, or V amplitudes exceed 4.5$\sigma_{\rm V}$ (dashed).}
\label{fig8}
\end{figure}

\section{Average magnetic parameters in the quiet Sun}

\begin{figure*}[!t]
\centering
\plotone{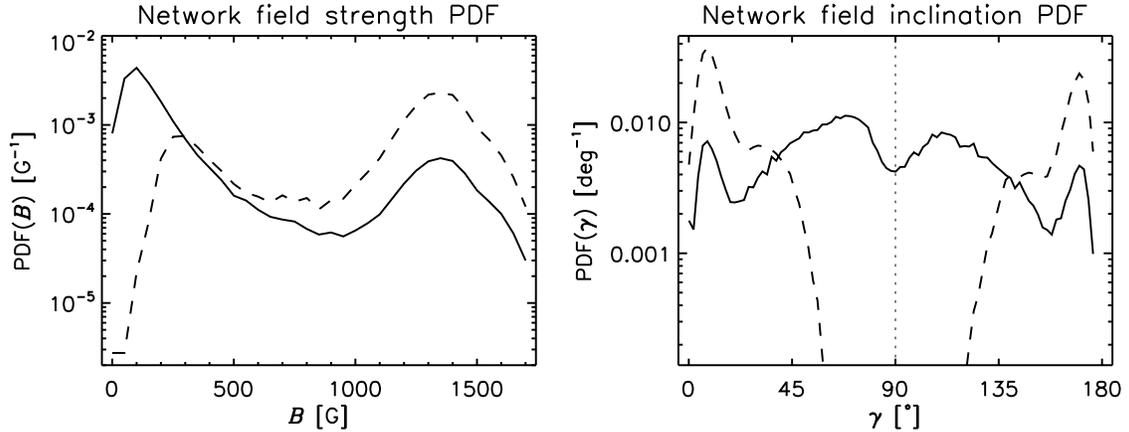}
\vspace*{1em}
\caption{Distribution of magnetic field strength (left) and inclination 
(right) in network regions resulting from the inversion of data set
\#1. The solid lines stand for all pixels within the selected network
areas and the dashed lines for pixels associated with network flux
tubes (see text).}
\label{fig9}
\end{figure*}

\subsection{The average field strengths}

We calculated the mean magnetic field strength, $\langle B\rangle
= \sum_{i=1}^NB_{i}/N$, and the mean vertical and horizontal
components of the field, $\langle B_{\rm z} \rangle =
\sum_{i=1}^N|B_{{\rm z},i}|/N = \sum_{i=1}^N|B_{i}\cos\gamma|/N  $ and $\langle B_{\rm h}
\rangle = \sum_{i=1}^N(B_{{\rm x},i}^2+B_{{\rm
y},i}^2)^{1/2}/N = \sum_{i=1}^N|B_{i}\sin\gamma|/N$, where $N$ is the
number of pixels, using the results from the analysis of the pixels
with Stokes Q or U signals above $4.5\sigma_{\rm Q,U}$ in the high S/N
time series. The values of $\langle B \rangle$, $\langle B_{\rm z}
\rangle$, and $\langle B_{\rm h} \rangle$ we obtain are 220, 64, and
198 G, respectively.  The dominance of $\langle B_{\rm h}
\rangle$ over $\langle B_{\rm z} \rangle$ indicates that the fields
are highly inclined in the IN, as first pointed out by
\cite{Lites1,Lites2}.  Since in the rest of pixels the linear
polarization signal does not surpass the noise threshold, they likely
have weaker fields. For that reason, the above values can be
interpreted as \emph{upper limits} for the mean field strength in the
internetwork. Note that the quantities $\langle B \rangle$, $\langle
B_{\rm z} \rangle$ and $\langle B_{\rm h} \rangle$ are independent of
the magnetic filling factor.

The ratio between the horizontal and vertical components of the field
in IN regions is $r \sim 3.1$ according to our
results. \cite{Lites1,Lites2} estimated a larger ratio $r\sim5$ for
all pixels within the field of view, but this value cannot be directly
compared to ours because it is based on ``apparent'' magnetic flux
densities rather than on intrinsic field strengths. In addition, the
method of \cite{Lites1,Lites2} uses less information than Stokes
inversions (for instance, Stokes I was not considered) and might be
affected by noise differently.

Our results partially agree with the MHD simulations of 
\cite{2008ApJ...680L..85S}, in which the magnetic field dynamics 
is mainly driven by flux expulsion and overshooting
convection. \cite{2008ApJ...680L..85S} computed the horizontally and
temporally averaged absolute vertical and horizontal magnetic field
components as functions of height for two different simulation runs
and found a maximum horizontal/vertical field component ratio of about
2.5 at 500~km (see Figure~1 in their paper). This is comparable
with the value we have obtained from the inversions.  However, the
average horizontal and vertical magnetic field strengths they find are
still smaller than ours.

Recently, \cite{2010A&A...513A...1D} have presented a set of
local dynamo simulations and have compared them with the results of
\citet{Lites1,Lites2} (see also \citealt{2009A&A...494.1091B} for a 
comparison between ground-based observations and
simulations). Following a similar approach as
\cite{2008ApJ...680L..85S}, these authors synthesized the Stokes
profiles of the 630~nm lines from the simulated models and degraded
them to the resolution of the {\it Hinode} SP
\citep{2008A&A...484L..17D}. After adding noise, they calculated the
longitudinal and transverse apparent flux densities. Their results
suggest that current local dynamo simulations explain the value of $r
\sim 5$ obtained by
\cite{Lites1,Lites2}. However, to reproduce the amount of transverse
and longitudinal flux and the variation of the flux ratio with
heliocentric angle,
\cite{2010A&A...513A...1D} had to artificially increase the average
magnetic field strength in the simulation by a factor of 2 or 3,
depending on the dynamo run.  With a factor of 3, their average field
turns out to be about 170~G at $\log \tau = 0$, in agreement with
Hanle measurements \citep{2004Natur.430..326T}.  This prompted
\cite{2010A&A...513A...1D} to argue that current {\it Hinode}
observations can be well explained by local dynamo processes.

The local dynamo simulations with artificially increased fields are
roughly compatible with our results. The average field strength of
170~G worked out by \cite{2010A&A...513A...1D} is below the upper
limit of 220~G we have deduced. In their simulations the vertical and
horizontal components of the field are also smaller than those
reported in this work. Finally, the ratio of horizontal to vertical
field components varies between 2 and 4 in the range $-2 <
\log \tau < -1$ (cf.~\citealt{2008A&A...481L...5S}), which is 
compatible with our value $r \sim 3.1$

The two mechanisms put forward to explain the existence of very
inclined fields in the IN, namely convective overshoot
\citep{2008ApJ...680L..85S} and local dynamo action
\citep{2010A&A...513A...1D}, may operate simultaneously 
or may not occur in the real Sun. To distinguish between the different
possibilities it is necessary to perform detailed comparisons of the
magnetic field distributions predicted by the simulations and those
obtained from the observations. This would give much more information
than just a single parameter such as the ratio of horizontal to
vertical magnetic field components.

\begin{figure*}[!t]
\centering
\epsscale{1.2}
\plotone{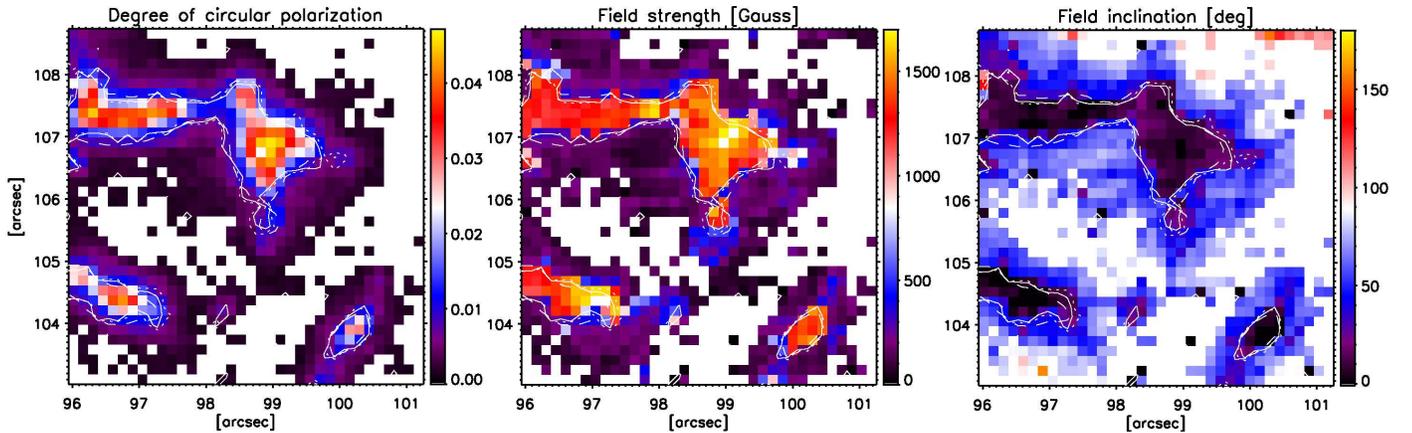}
\caption{Small area of about 5\arcsec $\times$5\arcsec\/ showing the
degree of circular polarization (left), the field strength (middle)
and the field inclination (right) resulting from the inversion of
normal map data. This area contains several network elements. The
white pixels were not inverted. Contour lines enclose areas with field
strengths greater than 700~G (dotted), inclinations lower than
$25\degree$ (solid), and polarization signal amplitudes above
$2\times10^{-2} \, I_{\rm QS}$ (dashed).}
\label{fig10}
\end{figure*}

\subsection{The average magnetic flux density}

The determination of the average flux density of IN regions has been
pursued for many years, which has resulted in more than forty papers
to date. Unfortunately, there is a large disparity between the values
obtained by the different authors. One of the reason is that the
estimates are strongly affected by the angular resolution of the
observations.  Also the different analysis techniques have contributed
to the discrepancies. As a result, the flux values reported in the
literature vary from the 2--3~Mx~cm$^{-2}$ of, e.g.,
\cite{1999ApJ...514..448L} and \cite{1994A&A...286..626K}, to 
the 15--20~Mx~cm$^{-2}$ found by \cite{2003dominguez},
\cite{2011A&A...530A..14V}, \cite{2009A&A...502..969B}, and
\cite{2008A&A...480..265B}. Studies using simultaneous observations of
visible and infrared lines give 11--15~Mx~cm$^{-2}$
\citep{khomenko}. An upper limit to the flux density seems to be
$\sim$~50~Mx~cm$^{-2}$ (e.g., \citealt{2001A&A...378..627F}).

We have calculated the average magnetic flux density $\langle
\phi \rangle= \sum_{i=1}^N|\phi_i|/N$ using the parameters
retrieved from the inversion of the high S/N time series. The flux of
individual pixels is computed as $\phi = f B \cos \gamma$. Note
that, as opposed to the magnetic field strength $B$, the flux density
depends on the magnetic filling factor.
The average is carried out in two different ways: taking into account
only those pixels with Stokes Q or U amplitudes above 4.5~$\sigma_{\rm
Q,U}$ and considering all pixels. In the latter case, the average flux
represents a \emph{lower limit} because we assign zero fluxes to
pixels without clear linear polarization signals (even if they show
large Stokes V amplitudes).

We find a flux density of 16.3~Mx~cm$^{-2}$ in the 27.4\% of the FOV
with significant linear polarization signals.  A lower limit for the
flux, considering all pixels, is 4.5~Mx~cm$^{-2}$. These results 
agree with the latest estimations of 11--16~Mx~cm$^{-2}$ by
\cite{Lites2011}. Our flux densities are slight smaller than those 
derived recently from the analysis of near infrared lines by
\cite{2009A&A...502..969B}, who found fluxes of about 26~G
and a lower limit of 22~G. They are also smaller than the ones
obtained from observations in the visible with the GREGOR Fabry-Perot
spectrometer attached to the German Vacuum Tower Telescope: according
to \cite{2008A&A...480..265B}, a lower limit for the flux would be
17~G. Finally, the net flux density, calculated as $\langle
\phi\rangle_{\mathrm{n}} = \sum_{i=1}^N\phi_i/N$, amounts to
$-2.9$~Mx~cm$^{-2}$.

\section{Differences with the network}

In this Section we show that the properties of network and
internetwork fields are rather different, as expected from their
different nature. The distribution of network field strengths and
inclinations obtained from the analysis of the normal map can be seen
in Figure~\ref{fig9}. The PDF for the field strength (solid line)
peaks at $\sim$~100~G but has a prominent hump centered at about
1.4~kG, in contrast to what is found in the IN where the majority of
fields are in the hG range (Figure~\ref{fig7}). In the same figure we
give the distribution of fields for the inner pixels of the network
flux concentrations (dashed line). They are selected by their circular
polarization amplitudes, which have to be larger than 0.03~I$_{\rm
QS}$. For those pixels, the peak at 100~G shifts toward stronger
fields and becomes smaller than the hump at 1.4~kG. The PDF vanishes
very rapidly below 300~G.

The inclination PDF corresponding to the network areas (solid line) is
dominated by inclined fields, but has secondary peaks at
$\sim$~10\degree\/ and 170\degree. The slightly asymmetric shape of the
distribution is due to the predominance of positive polarities among
the network areas selected for analysis.  When we consider only the
stronger network concentrations (dashed line), the peaks at
10\degree\/ and 170\degree\/ become dominant and the inclined fields
disappear.

The network PDFs show what is expected from intense field
concentrations, i.e., kG fields with predominantly vertical
orientations. The field strength and field inclination peaks we
observe are compatible with the results of analyses at lower spatial
resolution (e.g., \citealt{1987A&A...188..183S} and
\citealt{1994ApJ...424.1014S}). The large fraction of inclined
fields as well as the tail and hump of the field strength distribution
toward weak fields represent the contribution of network canopies to
the PDFs.  Network canopies (see e.g.\
\citealt{2010A&A...518A..50P} and references therein) are associated
with more inclined and weaker fields and can be observed at the
vicinity of strong network flux concentrations. To illustrate this,
Figure~\ref{fig10} displays the magnetic field strength and
inclination inferred from the ME inversion of a small area in data set
\#1 that includes network elements. Note that pixels corresponding to
the innermost part of the network elements show strong vertical
fields, while those at the periphery (canopy areas) are associated
with weak fields and moderate-to-large inclinations.

\section{Discussion and conclusions}
\label{Conclusions}

In this paper we have carried out a Milne-Eddington analysis of high
spatial resolution measurements of the quiet Sun taken with the {\it
Hinode} spectropolarimeter. To infer the magnetic field vector we have
applied the inversion strategy described by \cite{2007ApJ...662L..31O}
to two observations with different noise levels: $1.2 \times 10^{-3}$
and $3 \times 10^{-4}\, I_{\rm QS}$.

The magnetic field distributions obtained from the two data sets
are rather similar, suggesting that they are not biased by photon
noise.  A more detailed analysis reveals that differences exist
between the distributions when we consider the same type of
polarization signals (Stokes Q, U, or V larger than 0.5\%~I$_{\rm
QS}$) but different noise levels. We performed two tests to assess how
the noise and the calculation of the non-magnetic component affect the
inversion results. We reach the conclusion that noise is of little
concern.  Rather, the results may be influenced by the different
method we use to obtain the non-magnetic component in the two data
sets. This test suggest that the field strengths may be
slightly over-estimated in the high S/N data. The field inclinations
do not seem to depend on the exact way the non-magnetic component is
calculated.

We also provide magnetic field distributions in the IN based only on
pixels that show significant linear polarization signals
(Figure~\ref{fig7}). Most of them also have large circular
polarization signals. By restricting the analysis to those pixels we
make sure that the inversion results are accurate. The downside is
that they represent only 27.4\% of the surface area covered by the
IN. Our choice also implies that these high-precision magnetic field
distributions may be biased toward stronger and/or more inclined
fields---the ones that produce the larger linear polarization
signals. A better coverage of the IN requires more sensitive
measurements that are not yet available.

Keeping in mind these considerations, we discuss the main results of
our analysis in the following subsections.

\subsection{Internetwork fields are weak}

The inversion of {\it Hinode}/SP data presented here indicates that IN
fields are weak, at least in the 27.4\% of the FOV showing Stokes Q
or U signals well above the noise. This is in qualitative agreement
with the picture derived from the more magnetically sensitive
\ion{Fe}{1} lines at 1565~nm \citep{1995ApJ...446..421L,1999ApJ...514..448L,
2001ASPC..236..255C,2003A&A...408.1115K,
2006ApJ...646.1421D,2009A&A...502..969B} and with the simultaneous
inversion of the \ion{Fe}{1} 630~nm and 1565~nm lines performed by
\cite{2008A&A...477..953M}. Also \cite{2007A&A...466.1131R} and 
\cite{2009A&A...502..969B} found weak fields in the IN from 
the analysis of \ion{Fe}{1} 630~nm observations taken with the
Polarimetric Littrow Spectrograph \citep{2003AN....324..300S, 
2005A&A...437.1159B} at the German Vacuum Tower Telescope in Tenerife. 
Our findings are compatible with the results obtained by
\cite{2009ApJ...701.1032A} from the same data using 
Milne-Eddington inversions. 

There are other measurements that point to the existence of weak
fields in the IN. For example, the \ion{Mn}{1} 553~nm line analyzed by
\cite{2006A&A...454..663L} suggests that the IN is dominated by fields
below 600~G. Using the \ion{Mn}{1} line at 1526.2~nm,
\cite{2007ApJ...659..829A} found a Gaussian-shaped
distribution of field strengths centered at around 250-350~G. The
manganese lines are important because the hyperfine effects they show
make it possible to derive the strength of the field directly from the
{\em shape} of the profiles. Since the polarization amplitudes are not
used, the inferred field strengths are free from errors due to
uncertainties in the actual stray-light contamination or in the
magnetic filling factor.

The field strength PDF displayed in Figure~\ref{fig7} is not monotonic
and has a clear maximum in the region of weak fields. Although the
exact location of this maximum is still under debate, the peak itself
is most probably solar in origin. Several authors have argued that
peaks may result from the cancellation of magnetic flux at
sub-resolution scales (e.g., \citealt{2008A&A...477..953M}), but we do
not favor this interpretation because our spatial resolution is much
better than that of any previous measurement, which decreases the
possibility of cancellations. This does not mean that cancellations do
not exist at 0\farcs3, but that they are less frequent.

An important question is whether we should have considered more
complex magnetic topologies than those allowed by the Milne-Eddington
approximation. \cite{2011A&A...526A..60V} have recently analyzed
{\it Hinode}/SP data assuming that the magnetic fields of the IN are small
fibrils with sizes below $\sim$~100~km, the mean photon free path in
the solar photosphere. More specifically, they used a Micro-Structured
Magnetic Atmosphere (MISMA) of the type described by
\cite{1996ApJ...466..537S}.  With this model,
\cite{2011A&A...526A..60V} were able to fit the asymmetries of the
Stokes profiles recorded by {\it Hinode}
\citep{2011A&A...530A..14V}. They found that a broad range of 
field strengths (from hG to kG) are present in IN regions. In
particular, kG fields would dominate deep photospheric layers and hG
fields the layers above. They argued that an IN consisting of fibrils
with kG fields is partly compatible with the field strength PDFs shown
in Figure~\ref{fig4}, because the ME inferences carry information from
high photospheric layers. However, we doubt that our findings and
their results are truly compatible. First, the ME inferences cannot be
assigned to specific atmospheric layers but rather provide average
values over the region of formation of the spectral lines. Therefore,
our field strength distributions should be similar to the ones they
obtain, which is not the case. Second, the inversions performed by
\cite{2011A&A...526A..60V} are very dependent on the assumptions of the
MISMA model. In the absence of compelling proofs that the quiet Sun IN
consists of small-scale, mixed-polarity magnetic fibrils that are
still not resolved by {\it Hinode}, we tend to prefer more simplistic
approaches like the Milne-Eddington inversions described in this
paper.

Finally, we point out that our results seem to agree with those of 
\cite{2010A&A...517A..37S}. He reported the existence of two kinds of
magnetic structures in the quiet Sun, one characterized by strong and
vertical fields, and another by weak and isotropic fields.  We believe
that the strong fields are associated with the network elements, while
the weak ones are those found in the IN, although
\cite{2010A&A...517A..37S} cautioned that the association of IN fields
with the weak fields is not completely clear.

\begin{figure}[t]
\epsscale{1}
\begin{center}
\plotone{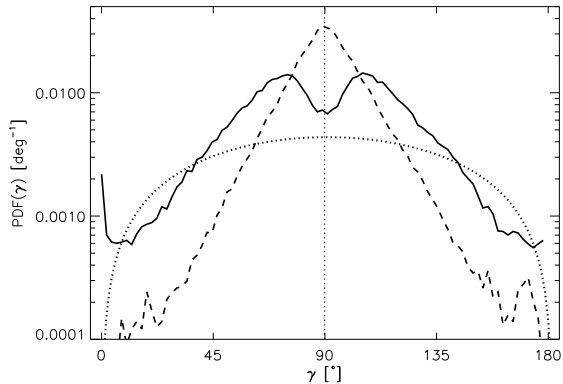}
\end{center} 
\caption{Magnetic field inclination PDF of IN regions in the 
high S/N time series where the linear {\em or} circular polarization
signals are larger than $1.35\times10^{-3}$~I$_{\rm QS}$
(corresponding to $4.5\sigma_{\rm V}$, solid line) and pixels where
the linear polarization exceeds the same threshold (dashed line). The
dotted line represents the distribution of inclinations for magnetic
field vectors with random orientation (isotropic case). Its shape is
given by a sine function, here represented in logarithmic scale.}
\label{fig11}
\end{figure}

\subsection{Internetwork fields are highly inclined}

The distribution of field inclinations deduced from the 27.4\% of the
FOV with clear Stokes Q or U signals in the high S/N time series
suggests that IN fields are very inclined. This result may be biased
by the requirement of significant linear polarization, but it is at
least valid for one quarter of the IN. The existence of nearly
horizontal hG fields in IN regions is compatible with the large
transverse magnetic fluxes found by \cite{Lites1,Lites2} and
\cite{2008A&A...477..953M}.  We caution, however, that ``very
inclined'' does not mean ``horizontal'': horizontal fields imply
inclinations of 90\degree\/, which are indeed the most abundant 
but not the only ones.

An aspect that is being discussed intensely in the literature is
whether or not the inclination PDF is compatible with an isotropic
distribution of magnetic field vectors in the FOV
(\citealt{2009ApJ...701.1032A, 2009A&A...506.1415B,
2010A&A...517A..37S}; see \citealt{2011ASPC..437..451S} for a review).
Based on the shape of the PDF, several authors have in
fact suggested that IN fields are ``turbulent''. Figure~\ref{fig11}
shows the distribution of magnetic field inclinations deduced from
pixels in the high S/N time series with circular or linear
polarization signals above 4.5 times the noise level, corresponding to
amplitudes of at least $1.35\times10^{-3}$~I$_{\rm QS}$ (solid line).
We also represent the PDFs of a random (i.e., isotropic) distribution
of magnetic fields (dotted line) and of pixels in the high S/N time
series that show
\emph{linear} polarization amplitudes above
$1.35\times10^{-3}$~I$_{\rm QS}$ (dashed line). As can be seen,
neither of the two distributions derived from the {\it Hinode} data
are compatible with an isotropic or quasi-isotropic distribution of
field vectors. The amount of very inclined fields obtained from the
inversions clearly exceeds that of a random distribution. Thus, we
conclude that IN fields are probably not isotropic, at least those
that show prominent linear polarization signals (representing 27.4\%
of the solar IN). A detailed analysis by Borrero \& Kobel (2011) also
points in the same direction.

\acknowledgments

We thank J.C.\ del Toro Iniesta, B.W.\ Lites, and J.M.\ Borrero for
stimulating discussions and comments that helped us to improve the
analysis. D.O.S. thanks the Japan Society for the Promotion of Science
(JSPS) for financial support through the postdoctoral fellowship
program for foreign researchers. This work has been partially funded
by the Spanish Ministerio de Innovaci\'on y Ciencia through projects
ESP2006-13030-C06-02 and PCI2006-A7-0624, and by Junta de
Andaluc\'{\i}a through Project P07-TEP-2687, including a percentage
from European FEDER funds. {\it Hinode} is a Japanese mission
developed and launched by ISAS/JAXA, collaborating with NAOJ as a
domestic partner, NASA and STFC (UK) as international
partners. Scientific operation of the {\it Hinode} mission is
conducted by the {\it Hinode} science team organized at
ISAS/JAXA. This team mainly consists of scientists from institutes in
the partner countries. Support for the post-launch operation is
provided by JAXA and NAOJ (Japan), STFC (UK), NASA, ESA, and NSC
(Norway). Use of NASA's Astrophysics Data System is gratefully
acknowledged.

\end{document}